\documentclass[twocolumn]{revtex4}

\usepackage{amsmath}
\usepackage{amssymb}
\usepackage{graphics}
\usepackage{tikz}
\usepackage{pgflibrarysnakes}

\newenvironment{definition}[1][Definition]{\begin{trivlist}
\item[\hskip \labelsep {\bfseries #1}]}{\end{trivlist}}

\newcommand{\bra}[1]{\left. \langle #1 \right|}
\newcommand{\ket}[1]{\left| #1 \rangle \right.}
\newcommand{\ketbra}[2]{\left| #1 \rangle \langle #2 \right|}
\newcommand{\scal}[2]{\langle #1 | #2 \rangle}

\begin{document}

\title{Hitting time for the continuous quantum walk}

\author{Martin Varbanov}
\email{varbanov@usc.edu}
\affiliation{Department of Physics and Astronomy, University of Southern California, \\
Los Angeles, CA  90089}

\author{Hari Krovi}
\email{krovi@nec-labs.com}
\affiliation{NEC Laboratories America, Inc., 4 Independence Way, suite 200, \\
Princeton, NJ  08540, USA}

\author{Todd A. Brun}
\email{tbrun@usc.edu}
\affiliation{Communication Sciences Institute, University of Southern California, \\
Los Angeles, CA  90089}

\date{\today}

\begin{abstract}
We define the hitting (or absorbing) time for the case of continuous quantum walks by measuring the walk at random times, according to a Poisson process with measurement rate $\lambda$. From this definition we derive an explicit formula for the hitting time, and explore its dependence on the measurement rate. As the measurement rate goes to either 0 or infinity the hitting time diverges; the first divergence reflects the weakness of the measurement, while the second limit results from the Quantum Zeno effect. Continuous-time quantum walks, like discrete-time quantum walks but unlike classical random walks, can have infinite hitting times.  We present several conditions for existence of infinite hitting times, and discuss the connection between infinite hitting times and graph symmetry.
\end{abstract}

\maketitle

\pagenumbering{arabic}
\section{Introduction}

There are two main types of quantum walks:  continuous-time and discrete-time quantum walks. Discrete-time quantum walks evolve by the application of a unitary evolution operator at discrete time intervals, and continuous-time walks evolve under a (usually time-independent) Hamiltonian. Continuous-time quantum walks have been defined by Farhi and Gutmann in \cite{Fa&Gu1} as a quantized version of continuous-time classical random walks. Classical random walks are used in computer science to design probabilistic algorithms for computational problems most notably for 3-satisfiability (3-SAT) \cite{MotRag95}. In a similar vein, quantum walks provide a framework for the design of quantum algorithms. As such quantum walks have been used in many quantum algorithms such as element distinctness \cite{Amb05}, matrix product verification \cite{BS06}, triangle finding \cite{MSS05} and group commutativity testing \cite{MN05}. Recently, a quantum algorithm for evaluating NAND trees has been proposed which uses a quantum walk as a part of the algorithm \cite{Fa&Go&Gu1}. In order to be able to understand how to better use quantum walks for algorithms, we need to study the properties of these walks. There have been many papers which study the behavior of quantum walks for various graphs. For example quantum walks on the line have been examined for the continuous-time case in  Refs.~\cite{Ch&Cl&etal1,Fa&Gu1,Ch&Fa&Gu1} and for the discrete-time case in \cite{Na&Vi1,BCGJW04,Br&Ca&Am1,Br&Ca&Am2,BCA03c}. The $N$-cycle is treated in \cite{AAKV00,TFMK}, and the hypercube in \cite{SKW03,MooRus02,JuKe2,Kr&Br2,Kr&Br1}. Quantum walks on general undirected graphs are defined in \cite{ViKe1,AnAl1}, and on directed graphs in \cite{Mon05}. Kendon \cite{Ken06} has a recent review of the work done in this field so far, focusing mainly on decoherence. Other reviews include an introductory review by Kempe in \cite{JuKe1}, and a review from the perspective of algorithms by Ambainis in \cite{AnAl1}.

Different quantitative characterizations of quantum walks are defined by analogy to the classical ones, such as mixing times, hitting (absorbing times), correlation times, etc. \cite{AAKV00}.  Often for this purpose the evolution of the quantum walk must be modified, to include not only the Hamiltonian evolution, but also a measurement process to extract information about the current state of the walk. There is a natural way to introduce such a measurement process in the discrete case:  namely, a measurement is made after each step of unitary evolution. The outcome from this measurement is used in defining the characteristic time scale in question. In the case of the continuous-time walk, such a natural definition of a measured walk does not exist.  There is no intrinsic time step after which we can perform the measurement.  Classically this is not a difficulty, because measurements do not disturb the state of the system.  The quantum case is quite different.  If we choose the measurement times arbitrarily, they can either be too long or too short with respect to the unitary evolution of the quantum walk. We can either miss important details in the evolution by measuring too infrequently, or overly distort the unitary evolution by measuring too often.  In the limiting case, we can completely freeze the evolution by the Quantum Zeno effect \cite{MiSu1}.

Hitting times for discrete-time quantum walks have been defined and analyzed in \cite{JuKe2,Kr&Br2}. The effect on mixing times of making random measurements, and its possible algorithmic applications, has been studied in \cite{KS04,Ric07,PeRi1}.  In this paper, we introduce a measurement process for the continuous-time quantum walk which gives rise to a definition and an analytical formula for the hitting time as function of the measurement rate (or equivalently, measurement strength). We explore the limits of measuring too weakly or too strongly, and show that the hitting time diverges in either case.  This suggests the existence of an {\it optimum} rate of measurement, which depends on the unitary dynamics of the particular walk.

We also show another difference from hitting times for classical random walks. In the classical case, a random walk on a finite connected graph always leads to finite hitting time for any vertex.  This is not true for the quantum case. The existence of infinite hitting times has been argued for discrete-time and continuous-time quantum walks in \cite{Kr&Br2}; in this paper we show this explicitly for the continuous-time quantum walks based on the definition of hitting time that we give, and derive conditions for the existence of infinite hitting times. Another sufficient condition that we prove is that if the complementary graph is not connected, this automatically leads to infinite hitting times for the continuous-time quantum walk on the original graph.

The paper is organized as follows. In section II, we describe how to introduce the measurement process, and derive formulas for the hitting time and probability. In section III, we give a condition for existence of infinite hitting times and prove that the hitting time diverges when the measurement rate goes to zero or infinity. In section IV, we give examples for the hitting times for certain graphs as a function of the measurement rate.  In section V, we give another sufficient condition for infinite hitting times.  A discussion follows in section VI.

\section{Hitting time definition for the continuous quantum walk}

We want to define hitting time for the continuous unitary evolution on undirected graph $\Gamma(V,E)$, where $V$ is the set of vertices and $E$ is the set of edges. Two vertices $v_1, v_2 \in V$ are connected if there exists an edge $e = \{v_1,v_2\} \in E$ (here $\{v_1,v_2\}$ should be taken as the {\it unordered} pair or set of the two vertices $v_1$ and $v_2$). Corresponding to the graph $\Gamma$, we assign the Hilbert space $\mathcal{H}_{\Gamma} = \ell^2(V)$. The vertex states in that Hilbert space are just labeled by the vertices of the graph---for $v \in V$, $\ket{v} \in \mathcal{H}$. They form an orthonormal basis for $\mathcal{H}_{\Gamma}$: $\scal{v_n}{v_m} = \delta_{nm}$.

The hitting time for classical random walks is defined naturally as the average time to find the walk in a specific vertex. When we turn to the quantum case the walk on the graph is not defined as a stochastic process on the vertices of the graph but as the unitary evolution of a closed quantum system with a Hilbert space defined as above. In order to be able to tell when the quantum walk has reached a vertex, we need to measure the system in order to gain information about the current state of the system. There are several reasonable ways to do that in the continuous-time case. We could perform strong measurements periodically with some fixed but arbitrary period $T$. This is not unlike in the discrete case, in which the period $T$ is given naturally by the walk itself:  a measurement is performed after each unitary evolution step. This way to perform a measurement in our case is unsatisfactory. We have no way to know how to choose $T$. If we choose it too small we could introduce too much decoherence, effectively masking the unitary evolution of the walk, or even worse, freezing it. If $T$ is too large then we can miss the moment when the walk actually reaches the final vertex.  And in general, the unitary transformation between measurements can be complicated and difficult to work out (unlike the discrete-time case).

Another way to measure the system is through strong measurements but performed at random times. The measurement times are chosen according to some probability distribution with some measurement rate. The advantage is that we don't introduce an artificial periodicity into the dynamics, and it allows one to calculate averaged effects over different measurement patterns.  The disadvantage is that it is still necessary to introduce a time scale for the measurements, this time given by the rate at which measurements are performed.

A third way to measure the system is using ``weak'' measurements, analogous to the way photodetection is described. In a small time period $\delta t$ we perform a measurement which either allows the system to evolve unitarily with a probability $1-\epsilon$, or performs a measurement to determine whether it has reached the final state with a probability $\epsilon$. In this case, the evolution is unitary for most of the time, with jumps at random time when the measurement is performed.  This case and the previous one are actually equivalent---the values of $\delta t$ and $\epsilon$ determine the measurement rate $\lambda$---but they give a somewhat different intuition about how to look at the measurement procedure. In the second case, it is natural to describe the random times is by a Poisson process with a given rate. In the third case the role of a rate is played by the strength of the ``weak'' measurement. In his work on mixing times \cite{PeRi1}, Peter Richter argued that the qualitative and even quantitative behavior of the system is not too sensitive to the exact details of the measurement scheme. This suggests that the first choice above might be as good as the the other two, but it still introduces a discrete structure which is not desirable in dealing with continuous evolution. In the following, we explore a measurement scheme described in the second and third cases.

We do a measurement to check whether the system is in the final state $\ket{v_f}$, given by the measurement operators $\{P_f, Q_f\}$ where $P_f = \ketbra{v_f}{v_f}$, $Q_f = I - P_f$. We have to specify the times when we perform the measurements. We will measure the system at random times distributed according to a Poisson process $X_t$ with rate $\lambda > 0$.  Each time we observe a jump in the Poisson process we measure the system. Between the moments at which we perform the measurements the system evolves unitarily with a Hamiltonian $H = - \gamma L$, where $L$ is the discrete version of the continuous Laplacian $\nabla^2$. In our case it is given by $L = A - D$, where $D$ is a diagonal matrix in the basis spanned by the vertex states with the degree of each vertex along the diagonal, and $A$ is the adjacency matrix of the graph \cite{Fa&Gu1,Ch&Go1}. In this paper we take $\gamma = 1$. If the degree of the vertex $v_n$ is $d_n$ then we have the following representation of $D$ and $A$
\begin{align}
D = \sum_n d_n \ketbra{v_n}{v_n}, \label{degreeMatrix} \\
A = \sum_{n,m} a_{nm} \ketbra{v_n}{v_m} ,  \label{adjacencyMatrix}
\end{align}
where
\begin{align}
a_{nm} =
    \begin{cases}
    1,      &\text{if } \{v_n, v_m\} \in E,\\
    0,      &\text{if } \{v_n, v_m\} \not\in E.
    \end{cases}
\end{align}
is the adjacency matrix of the graph $\Gamma(V,E)$.

Let $\omega = (t_1, t_2,...)$ be a sequence of random times when the jumps of the Poisson process are observed, with ($t_n \in \mathbb{R}, \ 0 < t_1 <t_2 < ...)$. For convenience we will take $t_0 = 0$. The sequences $\omega$ belong to a probability space $(\Omega, \mathfrak{F}, \mathbb{P})$ on which the Poisson process $X_t$ is defined:
\begin{align}
& X: \mathbb{R} \times \Omega \rightarrow \{0, 1, 2,... \}, \notag \\
& X_t(\omega) = n,\text{ if } t \in [t_n, t_{n+1}).
\end{align}
Here $\Omega$ is the set of all sequences of random times $\omega$, $\mathfrak{F}$ is the $\sigma$-algebra, generated by the Poisson process. The probability measure $\mathbb{P}$ on $\Omega$ is the one induced by the Poisson process. (For reference, see \cite{Gu&Bo1,AGu1}.)

For each sequence $\omega \in \Omega$ we define the hitting time as
\begin{equation}
\tau_{\omega} = \sum_{n=1}^{\infty} t_n p_n, \label{E1.1}
\end{equation}
where $p_n$ is the probability to find the system in the final state at time $t_n$ given that the system wasn't measured to be in the final state in any of the previous times $t_{n-1},..., t_1$. (For reference, see \cite{Kr&Br2,Ya&Ko&Im1}.) Define the intervals between jumps $t_{j-1}$ and $t_j$ as $\bar{t}_j = t_j - t_{j-1}$.  We want to average $\tau_{\omega}$ over all possible trajectories $\omega$ of the Poisson process, and take this as our definition for the hitting time:
\begin{equation}
\tau_h = \mathbb{E}_{\mathbb{P}} (\tau_{\omega}) = \int\limits_{\Omega} \tau_{\omega} d\mathbb{P}(\omega). \label{E1.2}
\end{equation}

We would like to find an analytical formula for $\tau_h$. From the definition of $p_n$, we have
\begin{align}
p_n = {\rm Tr} & \left\{ P_f \overleftarrow{\prod_{m=1}^{n-1}} \left( e^{-i(t_{m+1} - t_m)H} Q_f \right) e^{-it_1 H} \rho_i e^{it_1 H} \times \right. \notag\\
& \times \left. \overrightarrow{\prod_{m=1}^{n-1}} \left( Q_f e^{i(t_{m+1} -t_m) H} \right) \right\}.
\label{E1.14}
\end{align}
The arrow above the products signify whether the operators entering the products are ordered from left to right or vice versa, in other words $\overleftarrow{\prod_{m=1}^{n}} U_i = U_n U_{n-1} ... U_1$ and $\overrightarrow{\prod_{m=1}^{n}} U_i = U_1 U_2 ... U_n$.

We now introduce superoperators $\mathcal{U}_{\bar{t}}$ and $\mathcal{Q}_f$, defined by
\begin{align}
& \mathcal{U}_{\bar{t}} (X) = e^{-i \bar{t} H} X e^{i \bar{t} H}, \\
& \mathcal{Q}_f (X) = Q_f X Q_f,
\end{align}
and use them to rewrite \eqref{E1.14} as
\begin{equation}
p_n = {\rm Tr} \left\{ P_f \mathcal{U}_{\bar{t}_n} \circ \mathcal{Q}_f \circ \mathcal{U}_{\bar{t}_{n-1}} \circ \mathcal{Q}_f \circ ... \circ \mathcal{U}_{\bar{t}_1} (\rho_i) \right\}.
\end{equation}
We want to express the sum, \eqref{E1.1}, as a function of the $\{\bar{t}_n\}$. We do that by adding, subtracting and rearranging terms in the sum, assuming that it is absolutely convergent:
\begin{align}
\tau_{\omega} & = t_1 p_1 + t_2 p_2 + t_3 p_3 + t_4 p_4 + ... \notag \\
& = t_1 p_1 + t_2 p_2 - t_1 p_2 + t_1 p_2 + t_3 p_3 - t_2 p_3 \notag \\
& \quad + t_2 p_3 - t_1 p_3 + t_1 p_3 + t_4 p_4 - ... \notag \\
& = t_1(p_1 + p_2 + p_3 + ...) + (t_2 - t_1)(p_2 + p_3 + ...) \notag \\
& \quad + (t_3 - t_2)(p_3 + p_4 + ...) + ... \notag \\
& = \sum_{k=1}^{\infty} \bar{t}_k \sum_{n=k}^{\infty} p_n
\end{align}
Because the $\{t_n\}$ are the event times of a Poisson process, the interval times $\{\bar{t}_n\}$ are independent and identically random variables, exponentially distributed with parameter $\lambda$ and a probability density function given by
\begin{align}
f_{\bar{t}_n} (\bar{t}) =
    \begin{cases}
    \lambda e^{- \lambda \bar{t}},      &\bar{t} \geq 0\\
    0,                                  &\bar{t} < 0.
    \end{cases}
\end{align}
Knowing that, we reexpress formula \eqref{E1.2}:
\begin{equation}
\tau_h = \left( \prod_{l=1}^{\infty} \int\limits_0^{\infty} d\bar{t}_l \lambda e^{- \lambda \bar{t}_l} \right) (\tau_{\omega}).
\end{equation}
Then
\begin{equation}
\tau_h = \sum_{k=1}^{\infty} \sum_{n=k}^{\infty} \left( \prod_{l=1}^{\infty} \int\limits_0^{\infty} d\bar{t}_l \lambda e^{- \lambda \bar{t}_l} \right) (\bar{t}_k p_n).
\label{E1.5}
\end{equation}
In the above expression there are two types of integrals:
\begin{align}
A(X) & = \int\limits_0^{\infty} d\bar{t} \lambda e^{- \lambda \bar{t}} \mathcal{U}_{\bar{t}} (X), \\
B(X) & = \int\limits_0^{\infty} d\bar{t} \lambda e^{- \lambda \bar{t}} \bar{t} \mathcal{U}_{\bar{t}} (X).
\end{align}
Integrating by parts we get the following equations for the operators $A$ and $B$
\begin{align}
& A +\frac{i}{\lambda} [H,A] = X, \label{E1.3} \\
& B +\frac{i}{\lambda} [H,B] = \frac{1}{\lambda} A(X), \label{E1.4}
\end{align}
where $A(X)$ is the solution to the first equation.  (We will prove that this solution exists below.) Defining the superoperator
\[
\mathcal{L}_{\lambda}(X) = X + \frac{i}{\lambda} [H,X] ,
\]
we rewrite these equations as
\begin{align}
\mathcal{L}_{\lambda}(A) & = X, \\
\mathcal{L}_{\lambda}(B) & = \lambda^{-1} A(X).
\end{align}
We want to prove that the superoperator $\mathcal{L}_{\lambda}$ is invertible when $\lambda$ is a real number. For this we need to know how the adjoint of a superoperator is defined with respect to the Hilbert-Schmidt inner product for operators,
\begin{equation}
\langle X,Y \rangle_{HS} = {\rm Tr}(X^{\dagger} Y). \label{E1.17}
\end{equation}
Using this inner product, we see that if
\[
\mathcal{C}(X) = \sum_{n} c_n C_n X D_n^{\dagger} ,
\]
the adjoint of $\mathcal{C}(X)$ is given by
\[
\mathcal{C}^{\dagger}(X) = \sum_{n} c_n^* C_n^{\dagger} X D_n .
\]
From that it follows that $\mathcal{L}$ is a normal superoperator:
\begin{equation}
\mathcal{L}_{\lambda}^{\dagger} \circ \mathcal{L}_{\lambda} - \mathcal{L}_{\lambda} \circ \mathcal{L}_{\lambda}^{\dagger} = 0.
\end{equation}
This means that $\mathcal{L}_{\lambda}$ is diagonalizable. If $X_n$ is an eigenvector of $\mathcal{L}_{\lambda}$, then $X_n$ is an eigenvector of the hermitian and anti-hermitian parts of $\mathcal{L}_{\lambda}$ separately, which are given by
\[
\mathcal{L}_{\lambda}^H (X) = \frac{1}{2} (\mathcal{L}_{\lambda} + \mathcal{L}_{\lambda}^{\dagger})(X) = \mathcal{I} (X) = X
\]
and
\[
\mathcal{L}_{\lambda}^A (X) = \frac{1}{2} (\mathcal{L}_{\lambda} - \mathcal{L}_{\lambda}^{\dagger})(X) = \frac{i}{\lambda} [H,X],
\]
respectively. Let us denote the eigenvalue of $\mathcal{L}_{\lambda}^A$ corresponding to $X_n$ by $i x_n$ ($x_n \in \mathbb{R}$ because $\mathcal{L}_{\lambda}^A$ is anti-hermitian). Then
$\mathcal{L}_{\lambda} (X_n) = (1 + i x_n) X_n \neq 0$. This proves that each eigenvalue of $\mathcal{L}_{\lambda}$ is nonzero, and that $\mathcal{L}_{\lambda}$ is invertible.

The solutions to equations \eqref{E1.3} and \eqref{E1.4} are
\begin{align}
A & = \mathcal{L}_{\lambda}^{-1} (X), \\
B & = \lambda^{-1} \mathcal{L}_{\lambda}^{-2} (X).
\end{align}
Substituting these in \eqref{E1.5} we get
\begin{align}
\tau_h = \sum_{k=1}^{\infty} \sum_{n=k}^{\infty} \lambda^{-1} {\rm Tr} \left\{ P_f \left( \mathcal{L}_{\lambda}^{-1} \circ \mathcal{Q}_f \right)^{n-k} \circ \mathcal{L}_{\lambda}^{-2} \circ \right. \notag \\
\left. \circ \left( \mathcal{Q}_f \circ \mathcal{L}_{\lambda}^{-1} \right)^{k-1} (\rho_i) \right\}  \notag \\
= \sum_{k=1}^{\infty} \sum_{n=k}^{\infty} \lambda^{-1} {\rm Tr} \left\{ P_f \mathcal{L}_{\lambda}^{-1} \circ \left( \mathcal{Q}_f \circ \mathcal{L}_{\lambda}^{-1} \right)^{n-k} \circ \right. \notag \\
\left. \circ \mathcal{L}_{\lambda}^{-1} \circ \left( \mathcal{Q}_f \circ \mathcal{L}_{\lambda}^{-1} \right)^{k-1} (\rho_i) \right\}  \notag \\
= \sum_{k=0}^{\infty} \sum_{l=0}^{\infty} \lambda^{-1} {\rm Tr} \left\{ P_f \mathcal{L}_{\lambda}^{-1} \circ \left( \mathcal{Q}_f \circ \mathcal{L}_{\lambda}^{-1} \right)^{l} \circ \right. \notag \\
\left. \circ \mathcal{L}_{\lambda}^{-1} \circ \left( \mathcal{Q}_f \circ \mathcal{L}_{\lambda}^{-1} \right)^{k} (\rho_i) \right\}. \label{E1.13}
\end{align}
Note that the eigenvalues of $\mathcal{L}_{\lambda}$ are always greater than or equal to $1$ in absolute value, from which it follows that the eigenvalues of $\mathcal{Q}_f \circ \mathcal{L}_{\lambda}^{-1}$ are all less than or equal to $1$ in absolute value. If all the eigenvalues are strictly less than 1 in absolute value, the following sum exists:
\[
\sum_{k=0}^{\infty} \left( \mathcal{Q}_f \circ \mathcal{L}_{\lambda}^{-1} \right)^{l}
= \left( \mathcal{I} - \mathcal{Q}_f \circ \mathcal{L}_{\lambda}^{-1} \right)^{-1} .
\]
Substituting this into \eqref{E1.13} and denoting $\mathcal{N}_{\lambda} = \mathcal{L}_{\lambda} - \mathcal{Q}_f$, we get the following formula for the hitting time:
\begin{equation}
\tau_h = \lambda^{-1} {\rm Tr} \left\{ P_f \mathcal{N}_{\lambda}^{-2} (\rho_i) \right\}. \label{E1.8}
\end{equation}
This formula is closely analogous to the formula for the hitting time derived in \cite{Kr&Br1,Kr&Br2} for the case of a discrete-time quantum walk. If $\mathcal{Q}_f \circ \mathcal{L}_{\lambda}^{-1}$ has any eigenvalues equal to $1$, the inverse in the above formula should be thought of as a pseudoinverse.

Another quantity that may be defined is the total probability to ever hit the final vertex \cite{Ya&Ko&Im1,Kr&Br1}:
\begin{equation}
p_h = \sum_{n=1}^{\infty} p_n.
\end{equation}
We can derive a formula similar to \eqref{E1.8} for $p_h$:
\begin{equation}
p_h = {\rm Tr} \left\{ P_f \left( \mathcal{L}_{\lambda} - \mathcal{Q}_f \right)^{-1} (\rho_i) \right\}. \label{E1.16}
\end{equation}
When the hitting time is not infinite, or equivalently when the superoperator $\mathcal{L}_{\lambda} - \mathcal{Q}_f$ is invertible, $p_h = 1$. In the case of infinite hitting time, we replace the inverse with a pseudoinverse, as before.

There is another way to derive formula \eqref{E1.8}:  by looking at this procedure as an iterated weak measurement (case 3 that we discussed above). Instead of summing over all trajectories of the Poisson process at each time period $\delta t$, we perform a generalized measurement with measurement operators
\begin{align}
M_0 & = \sqrt{1 - \varepsilon^2} e^{-i \delta t H},\notag\\
M_1 & = \varepsilon P_f,\\
M_2 & = \varepsilon Q_f. \notag
\end{align}
These operators form a complete measurement as $\sum_{i=0}^{2} M_i^{\dagger} M_i = I$. The measurement is weak (in the particular sense of giving little information about the system on average) when $\varepsilon \ll 1$. Let us define a positive matrix $\rho^{c}$ describing the state of the system at time $t$ conditioned on the assumption that outcome ``2'' has not occurred up to this time. We measure the system repeatedly at intervals of time $\delta t$, using the same measurement operators, and if we don't observe outcome ``2'' the state of the system is described by the following matrix:
\begin{equation}
\rho^{c} (t + \delta t) = \sum_{i=0}^{1} M_i \rho^{c} (t) M_i^{\dagger}.
\end{equation}
We expand in powers of the small parameter $\varepsilon$ and take the limit $\varepsilon\rightarrow0$ and $\delta t\rightarrow0$, keeping the ratio $\varepsilon/\sqrt{\delta t}$ constant, and obtain a master equation for $\rho^{c} (t)$. This gives the connection between the strength of the measurement $\varepsilon$ and the measurement rate $\lambda = \lim_{\delta t \rightarrow 0} \varepsilon^2/\delta t$. After we expand to second order of $\varepsilon$ we get
\begin{align}
\rho^{c} (t + \delta t) = & (1 - \varepsilon^2) (1 - i \delta t H) \rho^{c} (1 + i \delta t H) \notag\\
& + \varepsilon^2 Q_f \rho^{c} Q_f + O(\varepsilon^3) \notag\\
= & \rho^{c} (t) - i \delta t [H,\rho^{c}] \notag\\
& - \varepsilon^2 \left( \rho^{c} (t) - Q_f \rho^{c} Q_f \right) + O(\varepsilon^3).
\end{align}
Taking the limit $\delta t \rightarrow 0$ and using the fact that
\begin{align}
& \lim_{\delta t \rightarrow 0} \frac{\rho^{c} (t + \delta t) - \rho^{c} (t)}{\delta t} = \frac{d \rho^{c}}{dt},\notag \\
& \lim_{\delta t \rightarrow 0} \frac{\varepsilon^2}{\delta t} = \lambda \notag
\end{align}
we arrive at the master equation for $\rho^{c}$:
\begin{equation}
\frac{d \rho^{c}}{dt} = - i [H,\rho^{c}] - \lambda \left(\rho^{c} - Q_f \rho^{c} Q_f \right) = - \lambda \mathcal{N}_{\lambda} (\rho^{c}). \label{E1.18}
\end{equation}
Note that $\rho^{c}$ is positive, but not normalized; the trace of $\rho^{c}$ is the probability that measurement result ``2'' has not been seen up until time $t$.
The total probability to hit the final vertex $p_h$ and the hitting time $\tau_h$ are given by:
\begin{align}
p_h = \lambda \int\limits_{0}^{\infty} {\rm Tr}\{ P_f \rho^{c} (t) \} dt,\label{E1.19}\\
\tau_h = \lambda \int\limits_{0}^{\infty} t {\rm Tr}\{ P_f \rho^{c} (t) \} dt. \label{E1.20}
\end{align}
Substituting the solution of \eqref{E1.18}
\begin{align}
\rho^{c} (t) = e^{- \lambda t \mathcal{N}_{\lambda}} (\rho^{c} (0)). \notag
\end{align}
in \eqref{E1.19} and \eqref{E1.20} and integrating by parts we obtain formulas \eqref{E1.8} and \eqref{E1.16} for the total probability to hit and the hitting time.

\section{Conditions for existence of infinite hitting times}

We want to prove that the existence of infinite hitting times is equivalent to the non-invertibility of the superoperator $\mathcal{L}_{\lambda} - \mathcal{Q}_f$. We will need the following definitions \cite{FeGa1,SeGo1}.

\begin{definition}{\textbf{1.}}
A matrix pencil $A + s B$ (where $A$ and $B$ are $n\times n$ matrices and $s$ is a complex number) is said to be {\it regular} if there exists at least one complex $s$ for which the pencil is nonsingular.
\end{definition}

\begin{definition}{\textbf{2.}}
A complex number $\bar{s}$ is a {\it finite eigenvalue} of the regular matrix pencil $A + s B$ if $det(A + \bar{s} B) = 0$.
\end{definition}

\begin{definition}{\textbf{3.}}
The regular matrix pencil $A + s B$ is said to have an {\it infinite eigenvalue} if $B$ is a singular matrix.
\end{definition}

In discrete-time quantum walks, infinite hitting times were observed for specific graphs and initial states \cite{Kr&Br2,Kr&Br1}. This occurs when, starting from the initial state, the total probability to ever find the walk at the final vertex is less than 1. It has been argued in the papers above that infinite hitting times occur given that the graph has a symmetry group that leads to a degenerate Hamiltonian of the quantum walk. The symmetry of the group leads to splitting the Hilbert space into invariant subspaces under the action of the group, on each of which the group acts with one of its irreducible representations. The evolutionary operator for the quantum walk leaves these subspace invariant under its action because the symmetry group of the graph is necessarily the symmetry group for the Hamiltonian. Thus if the final state is in one invariant subspace, but the initial state does not lie entirely in the same subspace, there will be a nonzero probability to never hit the final state. For such a situation to occur the final state, which we always assume to be a vertex state, must lie entirely in one of those invariant subspace. A sufficient condition for that in the case of a discrete-time quantum walk on a regular graph is the presence of an irreducible representation of the symmetry group in the Hamiltonian with a dimension larger than the dimension of the coin space. For the continuous-time quantum walk the equivalent condition is the presence of an irreducible representation with dimension larger than one. As Abelian groups always have one-dimensional representations, we might expect that in order to have infinite hitting times we need to have a symmetry group that is not Abelian. We will show that this is not true. The symmetry of the graph can lead to infinite hitting times even if the Hamiltonian is not degenerate, as is the case when the symmetry group of the graph is Abelian.  Having a non-Abelian symmetry group is a sufficient, but not a necessary, condition.

Another condition for the existence of infinite hitting times is connected to the invertibility of the superoperator $\mathcal{N}_{\lambda}$ which enters formulas \eqref{E1.8} and \eqref{E1.16}. Consider all operators $X$ such that $[H, X] = 0$ and $P_f X = X P_f = 0$, and denote the projector on the linear subspace of all such operators by $\mathcal{P}$. We will prove that $\mathcal{P} \neq 0$ if and only if $\mathcal{N}_{\lambda} = \mathcal{L}_{\lambda} - \mathcal{Q}_f$ is not regular.

If $\mathcal{P} \neq 0$ then choose $\bar{X}$ such that $\mathcal{P}(\bar{X}) = \bar{X} \neq 0$. Then $[H,\bar{X}] = 0$ and from $P_f \bar{X} = \bar{X} P_f = 0$ follows that $\mathcal{Q}_f (\bar{X}) = \bar{X}$. Thus $\mathcal{N}_{\lambda} (\bar{X}) = 0$ which means that $\mathcal{N}_{\lambda}$ is singular for every $\lambda$ and thus not regular.

If $\mathcal{N}_{\lambda}$ is not regular then it is non-invertible for any $\lambda$. Let us fix $\lambda$ to be real and different from 0. There exist a $\bar{X} \neq 0$ such that $\mathcal{N}_{\lambda}(\bar{X}) = 0$. We have already proven that $\mathcal{L}_{\lambda}$ is invertible for real $\lambda$. Then
\begin{equation}
\mathcal{N}_{\lambda} (\bar{X}) = \mathcal{L}_{\lambda} \circ (\mathcal{I} - \mathcal{L}_{\lambda}^{-1} \circ \mathcal{Q}_f) (\bar{X}) = 0
\end{equation}
and thus
\begin{equation}
\mathcal{L}_{\lambda}^{-1} \circ \mathcal{Q}_f (\bar{X}) = \bar{X}.
\end{equation}
Taking into account that all eigenvalues of $\mathcal{L}_{\lambda}$ are greater or equal to $1$ in absolute value the above equality is true only if
\begin{align}
\mathcal{L}_{\lambda}^{-1} (\bar{X}) = \bar{X},\\
\mathcal{Q}_f (\bar{X}) = \bar{X},
\end{align}
which are equivalent to
\begin{align}
[H, \bar{X}] = 0,\\
P_f \bar{X} = \bar{X} P_f = 0.
\end{align}
This means that $\mathcal{P}$ is nonzero.

Let's explicitly give the form of the projective superoperator $\mathcal{P}$ in terms of the projectors on the eigenspaces of the Hamiltonian $H$. For this purpose we have to define the {\it intersection} operation $\cap$ for orthogonal projections. For any two orthogonal projection operators $P_1$ and $P_2$, $P_1 \cap P_2$ will denote the orthogonal projector on the subspace which is the intersection of the subspaces onto which $P_1$ and $P_2$ project. Let the Hamiltonian $H$ have the following decomposition:
\begin{align}
H = \sum_{i=1}^{r} E_i P_i, \notag
\end{align}
where $E_i$ are the eigenvalues of $H$ ($E_i \neq E_j$ for $i \neq j$), $P_i$ are the projectors onto the eigenspaces of $H$ corresponding to eigenvalues $E_i$.  Since $H$ is Hermitian, $P_i P_j = \delta_{ij} P_i$, and ${\rm Tr}{P_i} = d_i$ is the multiplicity of the $E_i$ eigenvalue. The projector $\mathcal{P}$ is then
\begin{equation}
\mathcal{P} (X) = \sum_{i=1}^{r} (P_i \cap Q_f) X (P_i \cap Q_f).
\end{equation}
From this form it is easy to see that $\mathcal{P}$ is a completely positive superoperator.  As such, if it is different from $0$, then there must exist a density matrix $\rho$ such that $\mathcal{P}(\rho) = \rho$.  If the walk begins in such a state $\rho$, it will never arrive at the final vertex.

Now we will prove that if $\mathcal{N}_{\lambda}$ is regular as a matrix pencil then all its eigenvalues lie on the imaginary axis. Let's assume that
$\mathcal{N}_{\lambda}$ is invertible for some $\lambda \neq 0$ and $\mathcal{N}_{\lambda_0}$ is non-invertible for some $\lambda_0 \neq 0$, $\lambda_0
\neq \lambda$. Then $\mathcal{N}_{\lambda}(X) \neq 0$ for all $X \neq 0$ and there exists $X_0 \neq 0$ such that $\mathcal{N}_{\lambda_0}(X_0) = 0$.
Then $(\mathcal{N}_{\lambda} - \mathcal{N}_{\lambda_0})(X_0) = i (1/\lambda - 1/\lambda_0) [H, X_0] \neq 0$ and therefore $[H, X_0] \neq 0$.
Analogously $(\lambda \mathcal{N}_{\lambda} - \lambda_0 \mathcal{N}_{\lambda_0})(X_0) = (\lambda - \lambda_0) (\mathcal{I} - \mathcal{Q}_f)(X_0) \neq
0$ therefore $(\mathcal{I} - \mathcal{Q}_f)(X_0) \neq 0$. As $\mathcal{I} - \mathcal{Q}_f$ is a projector, it follows that
\begin{equation}
\langle X_0, (\mathcal{I} - \mathcal{Q}_f)(X_0) \rangle_{HS} \neq 0.
\end{equation}
Taking into account that $\mathcal{I} - \mathcal{Q}_f$ and $\mathcal{H}(\cdot) = [H, \cdot]$ are both Hermitian superoperators, $\langle X_0,
(\mathcal{I} - \mathcal{Q}_f)(X_0) \rangle_{HS}$ and $\langle X_0, \mathcal{H}(X_0) \rangle_{HS}$ are both real numbers. Denoting $\bar{\lambda}_0^r =
Re(1/\lambda_0)$ and $\bar{\lambda}_0^i = Im(1/\lambda_0)$ we have
\begin{equation}
\langle X_0, (\mathcal{I} -  \mathcal{Q}_f - \bar{\lambda}_0^i \mathcal{H})(X_0) \rangle_{HS} + i \bar{\lambda}_0^r \langle X_0, \mathcal{H}(X_0)
\rangle_{HS} = 0.
\end{equation}
This equality is only possible if both the real and imaginary parts vanish, implying that $\bar{\lambda}_0^r = 0$ and hence $Re(\lambda_0) = 0$.

Using this result, we will now prove that if $\mathcal{N}_{\lambda}$ is a regular matrix pencil, and thus doesn't have infinite eigenvalues, then the hitting time $\tau_h$ behaves regularly as a function of $\lambda$ on the real line:  it won't diverge for any real $\lambda$ except when $\lambda$ goes to 0 or infinity.  Physically, this means that when we measure either very weakly or very strongly we never find the particle in the final vertex.  The first limit is easy to understand---if we never measure, we will never find the particle anywhere.  The second limit---$\lambda$ going to infinity---corresponds to the Quantum Zeno effect, in which the evolution of the system is restricted to a subspace orthogonal to the final vertex.

To prove this conclusion, we represent superoperators as matrices using the following isomorphism:
\begin{align}
\phi:\mathcal{C}(\cdot) = & \sum_n c_n C_n (\cdot) D_n^{\dagger} \notag\\
\longrightarrow & \phi(\mathcal{C}) = \mathbf{C} = \sum_n c_n C_n \otimes D_n^*.
\end{align}
Now we represent the superoperator pencil $\mathcal{N}_\lambda$ by the matrix pencil
\begin{align}
\mathbf{N}_{\lambda} = & \phi(\mathcal{N}_{\lambda}) \notag\\
& = I \otimes I - Q_f \otimes Q_f^* - \frac{i}{\lambda} (H \otimes I - I \otimes H^*).
\end{align}
By assumption this matrix pencil is regular.  Every regular matrix pencil $A + s B$ has the following canonical form:
\begin{align}
A + \mu B = T \{N^{(m_1)},&...,N^{(m_p)},\notag\\
& J^{(n_1)}(\mu_{n_1}),...,J^{(n_q)}(\mu_{n_q})\} S, \label{E1.10}
\end{align}
where $T$ and $S$ are invertible matrices, constant with respect to $\mu$, and $\{N^{(m_1)},...,N^{(m_p)},J^{(n_1)},...,J^{(n_q)}\}$ is a block-diagonal matrix where the blocks $N^{(m)}$ and $J^{(n)}$ are square matrices of order $m$ and $n$ respectively of the form
\begin{align}
N^{(m)} & = \left(
\begin{array}{cccccc}
1 & \mu & 0 & \ldots & 0 & 0 \\
0 & 1 & \mu & \ldots & 0 & 0 \\
0 & 0 & 1 & \ldots & 0 & 0 \\
\vdots & \vdots & \vdots & \ddots & \vdots & \vdots \\
0 & 0 & 0 & \ldots & 1 & \mu \\
0 & 0 & 0 & \ldots & 0 & 1
\end{array}
\right),\\
J^{(n)} (\mu_l) & = \left(
\begin{array}{cccccc}
\mu + \mu_l & 1 & 0 & \ldots & 0 & 0 \\
0 & \mu + \mu_l & 1 & \ldots & 0 & 0 \\
0 & 0 & \mu + \mu_l & \ldots & 0 & 0 \\
\vdots & \vdots & \vdots & \ddots & \vdots & \vdots \\
0 & 0 & 0 & \ldots & \mu + \mu_l & 1 \\
0 & 0 & 0 & \ldots & 0 & \mu + \mu_l
\end{array}
\right),
\end{align}
or more succinctly $N^{(m)} = I^{(k)} + \mu K^{(m)}$ and $J^{(n)} = (\mu + \mu_0) I^{(n)} + K^{(n)}$, where $I^{(m)}$ is the the identity matrix of order $m$ and $K^{(m)}$ is a $m \times m$ matrix with ``1s'' immediately above the diagonal and ``0s'' everywhere else. The $N^{(m)}$ blocks are present when the matrix pencil has infinite eigenvalues and the $J^{(n)}$ blocks correspond to finite eigenvalues.

We want to examine the behavior of the inverse of a regular matrix pencil when $\mu$ approaches one of its eigenvalues. Assume that $\mu_0$ is a finite
eigenvalue of the pencil and the corresponding $n \times n$ block to that eigenvalue is $J(\mu_0)$. The inverse of this block is given by
\begin{equation}
J^{-1}(\mu_0) = \sum_{j=0}^{n-1} \frac{(-K)^j}{(\mu + \mu_0)^{j+1}}. \label{E1.9}
\end{equation}
In the above $K^0 = I$ and $K = K^{(n)}$. The inverses of all blocks that do not correspond to the eigenvalue $\mu_0$ will have regular behavior when $\mu$ approaches $- \mu_0$. We want to examine the behavior of $N^{(n)}$ when $\mu$ goes to infinity. Analogously the inverse of this block is given by
\begin{equation}
N^{-1} = \sum_{j=0}^{n-1} (- \mu K)^j.
\end{equation}
The inverse of the blocks corresponding to finite eigenvalues will have regular behavior when $\mu$ approaches infinity.

As the matrix pencil
\begin{equation}
\bar{\mathbf{N}}_{\mu} = \mathbf{N}_{1/\mu} = I \otimes I - Q_f \otimes Q_f^* + i \mu (H \otimes I - I \otimes H^*)
\end{equation}
is regular and has both finite ($\mu = 0$) and infinite eigenvalues (because both matrices $I \otimes I - Q_f \otimes Q_f^*$ and $i (H \otimes I - I
\otimes H^*)$ are singular), both types of blocks $N^{(k)}$ and $J^{(l)}$ are present in its normal form. If we express formula \eqref{E1.8} in terms
of matrices and vectors with $\mu = 1/\lambda$ we get an analogous expression
\begin{equation}
\tau_h = \mu P_f^v \cdot \bar{\mathbf{N}}_{\mu}^{-2} \rho_i^v,
\end{equation}
where $P_f^v$ and $\rho_i^v$ are the vectorized versions of the matrices $P_f$ and $\rho_i$.

When $\mu$ goes to 0 we can see from formula \eqref{E1.9} with $\mu_0 = 0$ that the asymptotic behavior of $\tau_h$ is given by
\begin{equation}
\tau_h = \tau_h^r (\mu) + \frac{1}{\mu} P_f^v \cdot (S^{-1} P_0 T^{-1})^2 \rho_i^v + O\left(\frac{1}{\mu^2}\right),
\end{equation}
where $\tau_h^r (\mu)$ is a function which is regular in a neighborhood of $\mu = 0$, T and S are the invertible matrices in the canonical form
\eqref{E1.10} of the matrix pencil $\bar{\mathbf{N}}_{\mu}$ and $P_0$ is the projector on the eigenspace with eigenvalue $\mu = 0$. Now as long as
$P_f^v \cdot (S^{-1} P_0 T^{-1})^2 \rho_i^v \neq 0$, $\tau_h$ will go to infinity when $\mu$ goes to 0($\lambda$ going to infinity) no matter whether terms of higher order, $O\left(\frac{1}{\mu^2}\right)$, are present or not.

Analogously, when $\mu$ goes to infinity, $\tau_h$ has the asymptotic behavior
\begin{equation}
\tau_h = \mu P_f^v \cdot (S^{-1} P_{\infty} T^{-1})^2 \rho_i^v + O(\mu^2),
\end{equation}
where $P_{\infty}$ is the projector on the eigenspace with infinite eigenvalue. Here again as long as $P_f^v \cdot (S^{-1} P_{\infty} T^{-1})^2
\rho_i^v \neq 0$, $\tau_h$ will go to infinity when $\mu$ goes to infinity ($\lambda$ going to 0) no matter whether terms of higher order, $O(\mu^2)$, are present or not.

As we shall see in the next section the hitting time in the examples that we will give below has the following form as a function of $\lambda$:
\begin{equation}
\tau_h = \tau_{(1)} \lambda + \frac{\tau_{(-1)}}{\lambda},
\end{equation}
where the constants $\tau_{(1)}$ and $\tau_{(-1)}$ depend on the particular graph.

\section{Examples}

\begin{figure}
\begin{center}
\begin{tikzpicture}[style=thick]
\draw (0,0) circle (1pt) node[anchor=north] {$v_1$} -- (1,0) circle (1pt) node[anchor=north] {$v_2$};
\draw (0.5,-0.75) node {$K_2$};
\draw (2,0) circle (1pt) node[anchor=north] {$v_1$} -- (3,0) circle (1pt) node[anchor=north] {$v_2$} -- (4,0) circle (1pt) node[anchor=north] {$v_3$};
\draw (3,-0.75) node {$L_3$};
\draw (5,0) circle (1pt) node[anchor=east] {$v_1$} -- (5.866,-0.5) circle (1pt) node[anchor=west] {$v_2$} -- (5.866,0.5) circle (1pt) node[anchor=west] {$v_3$} -- (5,0);
\draw (5.5,-0.75) node {$K_3$};
\draw (0,-2) circle (1pt) node[anchor=north] {$v_1$} -- (1,-2) circle (1pt) node[anchor=north] {$v_2$} -- (2,-2) circle (1pt) node[anchor=north]
{$v_3$} -- (3,-2) circle (1pt) node[anchor=north] {$v_4$};
\draw (1.5,-2.75) node {$L_4$};
\draw (4,-2) circle (1pt) node[anchor=north] {$v_1$} -- (5,-2) circle (1pt) node[anchor=north] {$v_2$} -- (5.866,-1.5) circle (1pt) node[anchor=west]
{$v_3$} -- (5.866,-2.5) circle (1pt) node[anchor=west] {$v_4$} -- (5,-2) circle (1pt);
\draw (5,-2.75) node {$KL_{3,1}$};
\draw (2,-4) circle (1pt) node[anchor=east] {$v_1$} -- (3,-4) circle (1pt) node[anchor=west] {$v_2$} -- (3.5,-3.134) circle (1pt) node[anchor=west]
{$v_3$};
\draw (3,-4) -- (3.5,-4.866) circle (1pt) node[anchor=west] {$v_4$};
\draw (3,-5) node {$S_4$};
\end{tikzpicture}
\end{center}
\caption{\label{F1.0} Graph examples (with assigned vertex labels)}
\end{figure}
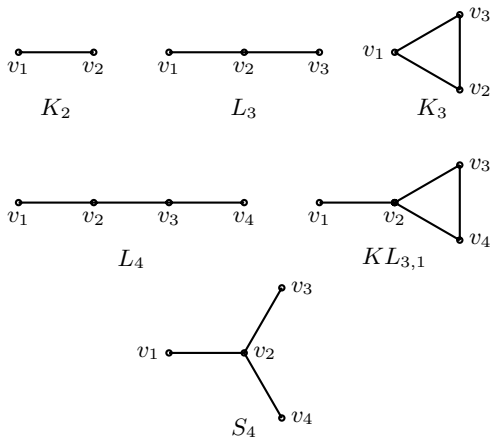

In this section, we will consider as examples the graphs in Fig.~\ref{F1.0}, using the labeling of the vertices given in the figure when necessary. We can see examples of infinite hitting times ($p_h \neq 0$) for the graphs $L_3$, $K_3$, $L_4$, $KL_{3,1}$ and $S_4$.

We can describe the probability to hit $p_h$ and the hitting time $\tau_h$ in another way, by specifying two operators $\mathfrak{P}_{\lambda}(\Gamma,v_f)$ and $\mathfrak{H}_{\lambda}(\Gamma,v_f)$, and calculating their expectations in the initial state:
\begin{align}
p_h = {\rm Tr}\{\mathfrak{P}_{\lambda}(\Gamma,v_f) \rho_i \},\\
\tau_h = {\rm Tr}\{\mathfrak{H}_{\lambda}(\Gamma,v_f) \rho_i \}.
\end{align}
Here $\rho_i$ is the density matrix describing the initial state of the system. These equations follow from formulas \eqref{E1.8} and \eqref{E1.16}, respectively. By using the definition of the Hilbert-Schmidt inner product, we derive the following formulas for $\mathfrak{P}_{\lambda}(\Gamma,v_f)$ and $\mathfrak{H}_{\lambda}(\Gamma,v_f)$:
\begin{align}
\mathfrak{P}_{\lambda}(\Gamma,v_f) = \left( \left( \mathcal{L}_{\lambda} - \mathcal{Q}_f \right)^{-1} \right)^{\dagger} (P_f),\\
\mathfrak{H}_{\lambda}(\Gamma,v_f) = \frac{1}{\lambda} \left( \left( \mathcal{L}_{\lambda} - \mathcal{Q}_f \right)^{-2} \right)^{\dagger} (P_f).
\end{align}

In the following, we will show the operators $\mathfrak{P}_{\lambda}(\Gamma,v_f)$ and $\mathfrak{H}_{\lambda}(\Gamma,v_f)$ in the vertex state basis for each of the graphs on Figure~\ref{F1.0}, and give a brief discussion of the quantum walk on each graph.  It is useful to describe the hitting probability and time in terms of these matrices, because they give the result for any starting state.

\subsubsection{Example 1}

The graph $K_2$ has the symmetry group $C_2$. As this group is Abelian, the Hamiltonian of this graph is nondegenerate. The two eigenvectors have nonzero overlap with both vertex states, and therefore there can be no infinite hitting times, as we can see from the matrices $\mathfrak{P}$ and $\mathfrak{H}$:
\begin{align}
\mathfrak{P}_{\lambda} (K_2,v_1) & = \left(
    \begin{array}{cc}
        1 & 0 \\
        0 & 1
    \end{array}
    \right)\\
\mathfrak{H}_{\lambda} (K_2,v_1) & = \left(
    \begin{array}{cc}
        \frac{2}{\lambda} & \frac{i}{\lambda} \\
        -\frac{i}{\lambda} & \frac{2}{\lambda} +\frac{\lambda}{2}
    \end{array}
    \right)
\end{align}
The dependence of the hitting time for vertex $v_1$ can include two terms:  the $2/\lambda$ term diverges as $\lambda\rightarrow0$, which simply represents the increasing time it takes to find the particle as the measurement rate goes to zero; if the system starts at vertex $v_2$ there is also a $\lambda/2$ term, which diverges as $\lambda\rightarrow\infty$ because of the Quantum Zeno effect:  as the measurement rate increases, we can ``freeze'' the system's evolution.

\subsubsection{Example 2}

The situation is different in the case of the $L_3$ graph. The graph again has symmetry group $C_2$, and the Hamiltonian has no degeneracies. Despite that, however, one of the three energy eigenstates has zero overlap with the $v_2$ vertex:  $(1/\sqrt2,0,-1/\sqrt2)$.  This means that even without degeneracy, there is an infinite hitting time for the final vertex $v_2$. This is not accidental and the symmetry of the graph is still responsible for the existence of this infinite hitting time. Under the action of $C_2$ each energy eigenstate $\ket{e_i}$ will have to be either symmetric or anti-symmetric. The Hilbert space thus splits into a symmetric and anti-symmetric subspaces which are orthogonal to each other. As the vertex state $\ket{v_2}$ is obviously symmetric under the action of the group it will be orthogonal to the anti-symmetric subspace. This is what leads to an infinite hitting time. This is a general observation for any graph that has $C_2$ as a symmetry group and vertex state that are left invariant under the action of the group. There are no infinite hitting times for reaching vertices $v_1$ and $v_3$. We can see all these properties by examining the matrices $\mathfrak{P}_{\lambda}$ and $\mathfrak{H}_{\lambda}$.
\begin{align}
\mathfrak{P}_{\lambda} (L_3,v_1) & =
    \left(
    \begin{array}{ccc}
    1 & 0 & 0 \\
    0 & 1 & 0 \\
    0 & 0 & 1
    \end{array}
    \right),\\
\mathfrak{H}_{\lambda} (L_3,v_1) & =
    \left(
    \begin{array}{ccc}
    \frac{3}{\lambda} & -\frac{1}{2 \lambda} + i & \frac{1}{2 \lambda} - \frac{i}{2} \\
    -\frac{1}{2 \lambda} - i & \lambda + \frac{4}{\lambda} & \frac{\lambda}{2} - \frac{1}{2 \lambda} + \frac{i}{2} \\
    \frac{1}{2 \lambda} + \frac{i}{2} & \frac{\lambda}{2} - \frac{1}{2 \lambda} - \frac{i}{2} & \frac{3 \lambda}{2} + \frac{3}{\lambda}
    \end{array}
    \right).
\end{align}
The existence of infinite hitting time for reaching $v_2$ can easily be seen from the $\mathfrak{P}$ matrix for $v_2$:
\begin{align}
\mathfrak{P}_{\lambda} (L_3,v_2) & =
    \left(
    \begin{array}{ccc}
    \frac{1}{2} & 0 & \frac{1}{2} \\
    0 & 1 & 0 \\
    \frac{1}{2} & 0 & \frac{1}{2}
    \end{array}
    \right), \label{E1.21}\\
\mathfrak{H}_{\lambda} (L_3,v_2) & =
    \left(
    \begin{array}{ccc}
    \frac{\lambda}{8} + \frac{9}{8 \lambda} & \frac{1}{4 \lambda} - \frac{i}{4} & \frac{\lambda}{8} + \frac{9}{8 \lambda} \\
    \frac{1}{4 \lambda} + \frac{i}{4} & \frac{2}{\lambda} & \frac{1}{4 \lambda} + \frac{i}{4} \\
    \frac{\lambda}{8} + \frac{9}{8 \lambda} & \frac{1}{4 \lambda} - \frac{i}{4} & \frac{\lambda}{8} + \frac{9}{8 \lambda}
    \end{array}
    \right). \label{E1.22}
\end{align}
As $\mathfrak{P}_{\lambda} (L_3,v_2)$ is not the identity, there must be initial states that will result in a less than 1 probability to hit $v_2$. For example, this will be true for any initial state which is a superposition of states $\ket{v_1}$ and $\ket{v_3}$.

\subsubsection{Example 3}

The graph $K_3$ has symmetry group $D_3$, and its Hamiltonian is degenerate.  It has infinite hitting times to hit any vertex.  If we calculate the $\mathfrak{P}$ and $\mathfrak{H}$ matrices for this graph, we discover a new property of these hitting time and hitting probability matrices.  The graph $K_3$ is not isomorphic to $L_3$, but its $\mathfrak{P}$ and $\mathfrak{H}$ matrices for any vertex of the $K_3$ are also given by \eqref{E1.21} and \eqref{E1.22} (or their appropriate cyclic permutations). This is because the Hamiltonians of the $L_3$ and $K_3$ graphs commute. We will observe the same kind of behavior below for other graphs with commuting Hamiltonians.

\subsubsection{Example 4}

The quantum walk on the graph $L_4$ has the same qualitative behavior as the walk on $L_2$. They both have the same symmetry group, $C_2$, as does the $L_3$ graph. But in the case of $L_4$, as in the case of $L_2$, there are no infinite hitting times.

\subsubsection{Examples 5 and 6}

We will examine the graphs $KL_{3,1}$ and $S_4$ together, because it turns out that their behavior is closely related.  The graph $KL_{3,1}$ again has $C_2$ for its symmetry group. The Hamiltonian is nondegenerate, but there are infinite hitting times for the vertices $v_1$ and $v_2$, due to the existence of an eigenvector which vanishes on those two vertices:  $(0,0,1/\sqrt2,-1/\sqrt2)$.  This is quite analogous to the case of the graph $L_3$. In general, graphs with $C_2$ symmetry will have infinite hitting times for hitting vertices that are fixed points under the action of the symmetry group.

The graph $S_4$ has $D_3$ as a symmetry group.  Its Hamiltonian {\it is} degenerate, and it has infinite hitting times to hit any vertex.  It turns out that the matrices $\mathfrak{P}$ and $\mathfrak{H}$ for hitting vertices $v_1$ and $v_2$ coincide with the same matrices for the graph $KL_{3,1}$:
\begin{align}
\mathfrak{P}_{\lambda}&(KL_{3,1},v_1) = \mathfrak{P}_{\lambda}(S_4,v_1) \notag\\
    &= \left(
    \begin{array}{cccc}
    1 & 0 & 0 & 0 \\
    0 & 1 & 0 & 0 \\
    0 & 0 & \frac{1}{2} & \frac{1}{2} \\
    0 & 0 & \frac{1}{2} & \frac{1}{2}
    \end{array}
    \right),\\
\mathfrak{H}_{\lambda}&(KL_{3,1},v_1) = \mathfrak{H}_{\lambda}(S_4,v_1) \notag\\
    &= \left(
    \begin{array}{cccc}
    \frac{3}{\lambda} & - \frac{1}{\lambda} + i & \frac{3}{4 \lambda} + \frac{i}{2} & \frac{3}{4 \lambda} + \frac{i}{2} \\
    - \frac{1}{\lambda} - i & \lambda + \frac{13}{2 \lambda} & \frac{\lambda}{2} - \frac{1}{\lambda} + \frac{i}{4} & \frac{\lambda}{2} -
    \frac{1}{\lambda} + \frac{i}{4} \\
    \frac{3}{4 \lambda} - \frac{i}{2} & \frac{\lambda}{2} - \frac{1}{\lambda} - \frac{i}{4} & \lambda + \frac{15}{8 \lambda} & \lambda + \frac{15}{8
    \lambda} \\
    \frac{3}{4 \lambda} - \frac{i}{2} & \frac{\lambda}{2} - \frac{1}{\lambda} - \frac{i}{4} & \lambda + \frac{15}{8 \lambda} & \lambda + \frac{15}{8
    \lambda}
    \end{array}
    \right),
\end{align}
\begin{align}
&\mathfrak{P}_{\lambda}(KL_{3,1},v_2) = \mathfrak{P}_{\lambda}(S_4,v_2) \notag\\
    &= \left(
    \begin{array}{cccc}
    \frac{1}{3} & 0 & \frac{1}{3} & \frac{1}{3} \\
    0 & 1 & 0 & 0 \\
    \frac{1}{3} & 0 & \frac{1}{3} & \frac{1}{3} \\
    \frac{1}{3} & 0 & \frac{1}{3} & \frac{1}{3}
    \end{array}
    \right),\\
&\mathfrak{H}_{\lambda}(KL_{3,1},v_2) = \mathfrak{H}_{\lambda}(S_4,v_2) \notag\\
    &= \left(
    \begin{array}{cccc}
    \frac{\lambda}{18} + \frac{8}{9 \lambda} & \frac{1}{3 \lambda} - \frac{i}{6} & \frac{\lambda}{18} + \frac{8}{9 \lambda} & \frac{\lambda}{18} +
    \frac{8}{9 \lambda} \\
    \frac{1}{3 \lambda} + \frac{i}{6} & \frac{2}{\lambda} & \frac{1}{3 \lambda} + \frac{i}{6} & \frac{1}{3 \lambda} + \frac{i}{6} \\
    \frac{\lambda}{18} + \frac{8}{9 \lambda} & \frac{1}{3 \lambda} - \frac{i}{6} & \frac{\lambda}{18} + \frac{8}{9 \lambda} & \frac{\lambda}{18} +
    \frac{8}{9 \lambda} \\
    \frac{\lambda}{18} + \frac{8}{9 \lambda} & \frac{1}{3 \lambda} - \frac{i}{6} & \frac{\lambda}{18} + \frac{8}{9 \lambda} & \frac{\lambda}{18} +
    \frac{8}{9 \lambda}
    \end{array}
    \right).
\end{align}
Just as with graphs $L_3$ and $K_3$, these two graphs have the same matrices because the Hamiltonians of the graphs $KL_{3,1}$ and $S_4$ commute; and, as we saw above, this produces similar dynamics when we measure the walk in the corresponding final vertices $v_1$ and $v_2$.

This is {\it not} the case, however, when the final vertex is $v_3$ or $v_4$ for these graphs.  For $S_4$, the $\mathfrak{P}$ and $\mathfrak{H}$ matrices for $v_3$ and $v_4$ can be obtained from those above by interchanging $v_1$ with $v_3$ or $v_4$.  For $KL_{3,1}$, however the matrices are
\begin{align}
&\mathfrak{P}_{\lambda}(KL_{3,1},v_3) \notag\\
    &= \left(
    \begin{array}{cccc}
    1 & 0 & 0 & 0 \\
    0 & 1 & 0 & 0 \\
    0 & 0 & 1 & 0 \\
    0 & 0 & 0 & 1
    \end{array}
    \right),\\
&\mathfrak{H}_{\lambda}(KL_{3,1},v_3) \notag\\
    &= \left(
    \begin{array}{cccc}
    \lambda + \frac{5}{\lambda} & - \frac{\lambda}{2} - \frac{1}{\lambda} - \frac{i}{2} & 0 & - \frac{\lambda}{2} - i \\
    - \frac{\lambda}{2} - \frac{1}{\lambda} + \frac{i}{2} & \frac{5}{2 \lambda} + \frac{7}{\lambda} & - \frac{1}{\lambda} - \frac{3 i}{2} & - \lambda -
    \frac{1}{\lambda} + \frac{i}{2} \\
     0 & - \frac{1}{\lambda} + \frac{3 i}{2} & \frac{4}{\lambda} & \frac{1}{\lambda} \\
    \frac{\lambda}{18} + \frac{8}{9 \lambda} & - \lambda - \frac{1}{\lambda} - \frac{i}{2} & \frac{1}{\lambda} & \lambda + \frac{4}{\lambda}
    \end{array}
    \right).
\end{align}
The matrices $\mathfrak{P}_{\lambda}(KL_{3,1},v_4)$ and $\mathfrak{H}_{\lambda}(KL_{3,1},v_4)$ can be found by interchanging $v_3$ and $v_4$ in the matrices above.

The infinite hitting times for the $L_3$ and $KL_{3,1}$ graphs can be understood to arise because the Hamiltonian of those graphs commutes with the Hamiltonian of a more symmetric graph. As we shall see in the next section, this fact can lead to infinite hitting times under certain circumstances.

\section{Infinite hitting times for graphs with non-connected complementary graph}

We will now look in a little more detail at infinite hitting times, which are one of the most surprising differences between classical random walks and quantum walks. In the classical case, if the graph is finite and connected, the probability to reach any vertex starting from any other is always 1. That is not the case for quantum walks. A sufficient condition for infinite hitting times was given in \cite{Kr&Br3}:  if the graph's Hamiltonian is sufficiently degenerate, infinite hitting times will always exist.  For continuous-time walks, any degeneracy at all is sufficient.  (This is a sufficient but not a necessary condition because, as we've shown above, even graphs with nondegenerate Hamiltonians may have infinite hitting times.)  We will now show that another sufficient condition for a continuous-time quantum walk to have an infinite hitting time is the non-connectedness of the {\it complementary} graph. Consider a graph $\Gamma$ with $n$ vertices and Hamiltonian $H_{\Gamma}$ given by the usual expressions Eq.~(\ref{degreeMatrix}) and Eq.~(\ref{adjacencyMatrix}). The complete graph $K_n$ with $n$ vertices has the following Hamiltonian (in the basis spanned by the vertex states):
\begin{equation}
H_{K_n} = \left(
\begin{array}{cccc}
n-1 & -1 & \ldots & -1 \\
-1 & n-1 & \ldots & -1 \\
\vdots & \vdots & \ddots & \vdots \\
-1 & -1 & \ldots & n - 1
\end{array}
\right). \label{E1.11}
\end{equation}
This can be rewritten more succinctly as
\begin{equation}
H_{K_n} = n \left( I - \ketbra{\psi_0}{\psi_0} \right) = n \bar{P_0},
\end{equation}
where $\ket{\psi_0} = \frac{1}{\sqrt{n}} \sum_{k=1}^n \ket{k}$ and $\bar{P_0} = I - \ketbra{\psi_0}{\psi_0}$.

The complementary graph $\Gamma^c$ of a graph $\Gamma$ is obtained by connecting vertices that are {\it not} connected in the original graph $\Gamma$, and removing the edges that {\it are} present in the original graph. Then it's easy to see that the Hamiltonian of $\Gamma^c$ is
\begin{equation}
H_{\Gamma^c} = H_{K_n} - H_{\Gamma}. \label{E1.12}
\end{equation}
Another observation is that the Hamiltonian of every graph $\Gamma$ commutes with the Hamiltonian of the complete graph with the same number of vertices:
\begin{equation}
[ H_{\Gamma} , H_{K_n} ] = 0. \label{E1.15}
\end{equation}
This follows from the observation that $\ket{\psi_0}$ is always an eigenvector of $H_{\Gamma}$ with eigenvalue 0. As
\begin{equation}
H_{\Gamma} = \bar{P_0} H_{\Gamma} \bar{P_0} ,
\end{equation}
\eqref{E1.15} is obvious. We can see from \eqref{E1.12} that $[ H_{\Gamma}, H_{\Gamma^c}] = 0$.

Let us assume that the graph $\Gamma$ is connected but the complementary graph $\Gamma^c$ is not, and consider the quantum walk on $\Gamma^c$. Because $\Gamma^c$ is not connected, there are initial states that never reach a particular final vertex if the initial state includes only vertices which are not connected to the final vertex. Let's consider an initial state $\ket{\psi_i}$ that contains only vertex states that belong to one of the connected components of $\Gamma^c$.  Let us further assume that
\[
\scal{\psi_0}{\psi_i} = \frac{1}{\sqrt{n}} \sum_{k=1}^n \scal{k}{\psi_i} = 0,
\]
which is always possible if the connected component has more than one vertex.  Since $\ket{\psi_i}$ is orthogonal to $\ket{\psi_0}$, it immediately follows that it is an eigenstate of $H_{K_n}$ with eigenvalue $n$.

If the final state  $\ket{\psi_f}$ contains only vertices belonging to a {\it different} connected component of $\Gamma^c$, the probability to ever reach the final state is 0:
\begin{equation}
\bra{\psi_f} e^{-i t H_{\Gamma^c}} \ket{\psi_i} = 0 \ \forall\, t.
\end{equation}
Putting all this together, we can see now that
\begin{align}
\bra{\psi_f} e^{-i t H_{\Gamma}} \ket{\psi_i} = \bra{\psi_f} e^{-i t (H_{K_n} - H_{\Gamma^c})} \ket{\psi_i}\nonumber\\
= e^{-i t n} \bra{\psi_f} e^{i t H_{\Gamma^c}} \ket{\psi_i} = 0.
\end{align}
This proves that existence of infinite hitting time for the original graph $\Gamma$.

We note that similar considerations may apply in some cases if the complete graph is replaced by a symmetric graph whose Hamiltonian commutes with with $H_{\Gamma}$. An example of this is the similarity of the dynamics of the $KL_{3,1}$ and $S_4$ graphs.

Finally, we note that this sufficient condition for infinite hitting times is not particularly strong.  For a graph with a large number of vertices, the complementary graph is almost always connected.

\section{Discussion}

We have examined continuous-time quantum walks and studied natural definitions for the hitting time. After considering different possibilities for introducing a measurement scheme, one of them emerges as a natural one for the continuous case:  measuring the presence or absence of the particle at the final vertex at Poisson-distributed random times, with an adjustable rate $\lambda$.  This is exactly equivalent to performing a particular type of weak measurement at frequent intervals, in the limit yielding continuous monitoring with time-resolution $1/\lambda$.  Using this measurement scheme, we derived an analytical formula for the hitting time which closely resembles the formula for the discrete-time case.

This formula enables us to find a necessary and sufficient condition for the existence of quantum walks with infinite hitting times, namely that a certain superoperator pencil is not regular.  In the case of finite hitting-times, the dependance of the hitting time on the rate of the measurement was studied, and the intuitive expectation for its behavior in the limits of weak and strong measurement rate was confirmed.  In particular, as the measurement rate goes to infinity, the hitting time can diverge due to the Quantum Zeno effect.

As in the discrete case, the symmetry of the graph plays a very strong role in the emergence of infinite hitting times. The graph symmetry group, if big enough, causes degeneracies in the eigenspectrum of the Hamiltonian which in turn leads to the emergence of infinite hitting times for certain vertices. But this is not the only way in which symmetry can lead to infinite hitting times. Even when no degeneracy is present, symmetry can cause some eigenvectors of the Hamiltonian to have zero overlap with some vertex states, as in the case of the $L_3$ and $KL_{3,1}$ graphs examined in section IV. This can be attributed to the fact that under the action of the group $C_2$, the Hilbert space splits into symmetric and antisymmetric subspaces, and some eigenvectors from the antisymmetric subspace could have zero overlap with certain vertex states. A further study exploring this idea is necessary to see if similar effects can occur for other symmetry groups.

Finally, in section V we show another condition for infinite hitting times. We have shown that the quantum walk on a connected graph can have infinite hitting times if the complementary graph is disconnected. This is in sharp contrast with the classical case, where every random walk on a connected graph will hit any vertex with probability 1 at long times. While this new condition is rather specific, it is possible that it can be generalized by replacing the completely-connected graph with some other highly symmetric graph such that the Hamiltonian still commutes with the Hamiltonian of the original graph. It may be possible to explain any infinite hitting times on any graph in this way, giving a unifying view of the whole subject.  It is clear that many questions remain, and that hitting times for continuous-time quantum walks are a very fruitful area of research.

\begin{acknowledgements}
TAB and HK acknowledge support from NSF Career Grant CCF-0448658. TAB, HK and MV acknowledge support from NSF Grant CCF-0524822. MV acknowledges support from NSF Grand CCF-0524811.
\end{acknowledgements}

\end{document}